\documentclass[twocolumn,prb,showpacs,amsmath,amssymb]{revtex4}
\usepackage{graphicx}
\usepackage{hyperref}
\begin{document}
\title{Theory of Non-Coherent Spin Pumps}
\author{Eran Sela and Yuval Oreg \date{\today}}
\affiliation{Department of Condensed Matter Physics, Weizmann
Institute of Science, Rehovot, 76100, ISRAEL }

\begin{abstract}
We study electron pumps in the absence of interference effects
paying attention to the spin degree of freedom. Electron-electron
exchange interactions combined with a variation of external
parameters, such as magnetic field and gate potentials, affect the
compressibility-spin-tensor whose components determine the
non-coherent parts of the charge and spin pumped currents. An
appropriate choice of the trajectory in the parameter space
generates an arbitrary ratio of spin to charge pumped currents.
After showing that the addition of {\it dephasing leads} to a full
quantum coherent system diminishes the interference contribution,
but leaves the non-coherent (classical) contribution intact, we
apply the theory of the classical term for several examples. We
show that when exchange interactions are included one can
construct a source of pure spin current, with a constant magnetic
field and a periodic variation of gate potentials only. We discuss
the possibility to observe it experimentally in GaAs
heterostructures.

\end{abstract}

 \maketitle

\section{Introduction}

\label{se:introduction}

A pump is a very common device, it appears in many shapes and
types, in physical systems as well as biological systems. A pump
converts a periodic variation of the parameters that control it
into a direct current. For example, in an Archimedes (or Auger)
screw pump after one revolution ``buckets" of water are lifted up,
while the blades of the screw return to their position at the
initial stage of the revolution. \cite{Rorres00}

As modern electronic devices, including electron pumps, become
smaller and get into the mesoscopic size of a few nanometer to a
fraction of micrometer, we need to consider also the wavy, i.e.,
quantum mechanical, nature of the electrons. Similar to the
classical devices, a quantum pump converts periodic variations of
its parameters at frequency $\omega$ into a DC current.
\cite{Thouless83}

In both quantum and classical devices the pumping rate is
proportional to the liquid pumped in one turn of the pump and to
the revolution rate $\omega$. When the Archimedean screw is
rotated too fast, turbulence and sloshing prevents the buckets
from being filled and the pump stops to operate. Similarly, in
small electronic devices, the pumping rate is proportional to
{$\omega$} only in the adiabatic limit --- when $\omega$ is small
enough.

While pumps exist for both interacting coherent quantum systems
and interacting classical systems, the main theoretical studies of
mesoscopic pumps were concentrated on the wavy nature of the
electrons describing the pumps in terms of non-interacting
coherent-scattering
theory.\cite{Spivak95,Brouwer98,Avron02,Entin02} Several works
discuss the effect of dephasing
\cite{Moskalets01,Cremers02,Brouwer02} which smears the wavy
nature of electrons, suppresses interference effects and renders
the system non-coherent, i.e.,~{\em classical}. The complexity of
the full quantum problem in presence of interactions\cite{Yigal94}
allowed its study only in few examples of open quantum
dots~\cite{Aleiner98a} and Luttinger liquid.\cite{Sharma01}

On the other hand, the early experimental studies of
quantum-dot-turnstile pump~\cite{Kouwenhoven91,Pothier91} are
described in classical terms of interacting systems, namely,
capacitance and oscillating ``resistances" of tunnel barriers. In
a more recent experiment,~\cite{Switkes99} coherent pumping was
observed. (However, parasitical rectification effects may be
relevant.\cite{Brouwer01a,Rect03})

In this manuscript we study the classical non-coherent effect of
pumping and in particular we show how this classical effect
directly emerges out of the quantum mechanical
formulation\cite{Brouwer98} when dephasing sources exist. While
the quantum-scattering description may include physics related to
the spin degree of freedom,\cite{Sharma03} it treats
electron-electron interaction on the Hartree level only. In this
manuscript we include both direct and exchange interaction within
the framework of a classical theory. This enables us to suggest a
scheme to build a pure ``spin battery" (see Sec.~\ref{se:hartree
fock}) which is a key concept in the field of spintronics.

The remainder of the paper is organized as follows: in
Sec.~\ref{se:Elec_circ} we derive an expression for the
non-coherent contribution for small electron pumps using a
classical model that neglects interference effects completely. The
formulation of pumps in classical terms allows us to include
rather easily the effects of interactions between the electrons.

To describe effects of dephasing  in a controlled
manner~\cite{Moskalets01,Cremers02}  we include in
Sec.~\ref{se:comparison}, following Ref.~\onlinecite{Moskalets01},
the effect of additional voltage leads on the (non-interacting
adiabatic) quantum scattering theory of pumps.
\cite{Spivak95,Brouwer98,Avron02,Entin02} We show that when the
coupling to the dephasing leads is tuned properly to cause
complete dephasing, the Brouwer formula,~\cite{Brouwer98} which
relates the pumping current to the scattering matrix and its time
derivatives, reduces to the classical expressions developed in
Sec.~\ref{se:Elec_circ}. We compare the magnitudes of the quantum
and classical contributions and study under what conditions the
classical contribution dominates.

After constructing and justifying the classical theory of
non-coherent pumps, we generalize it in Sec.~\ref{se:spin pola} to
deal with spinfull electrons endowing our result with a
topological interpretation.
Finally, in Sec.~\ref{se:2DEG} we deal with a possible realization
of spin pumps in two dimensional electron gas of GaAs
heterostructures. We show that under rather general conditions the
spin current (similar to the Einstein`s DC conductance formula) is
given in terms of the thermodynamic density of states tensor of
the system (the compressibility tensor). In the presence of a
constant magnetic field, we find an appropriate trajectory in the
parameter space such that a \emph{pure} spin current will flow.
This effect vanishes in the absence of exchange interaction.

In appendix~\ref{se:pumpvsbiased} we discuss the relation between
the classical non-coherent pumped current and the biased current,
generated by rectification effects.\cite{Rect03} In
appendix~\ref{se:linearresponse} we explore the relation between
our theory and the theory of non-linear response. In this
manuscript we do not include effects of charge discreteness
leaving this for a future study.\cite{Sela04}

\section{Classical Description of non-coherent Pumps}
\label{se:Elec_circ}
\begin{figure}[h]
\begin{center}
\includegraphics*[width=80mm]{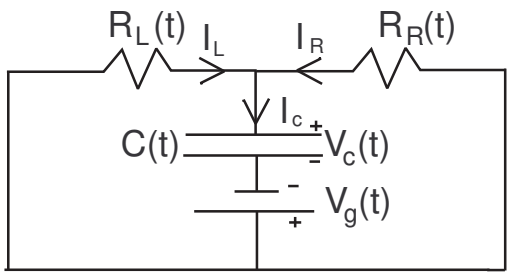}
\caption{Equivalent circuit of a classical pump. A typical pumping
cycle occurs when we charge the capacitor through the left
resistor (keeping $R_R \gg R_L$) and discharge it through the
right resistor (when $R_R \ll R_L$). In that case  a net DC
current flows in the clockwise direction in the large current
loop.
\label{fg:circuit}}
\end{center}
\end{figure}

Consider the electrical circuit depicted in Fig.~\ref{fg:circuit}.
Its purpose is to charge the capacitor $C$ from left and then to
discharge it to the right, thus producing a net current from left
to right without a bias voltage. The gate voltage, $V_g$,
periodically charges and discharges the capacitor while the
resistors, $R_L$ and $R_R$, control the direction of the charging
and discharging processes.

To analyze the system it is convenient to define an asymmetry
parameter
\begin{equation}
\label{eq:alpha} \alpha(R_L,R_R)=\frac{1}{2}
\frac{R_R-R_L}{R_R+R_L}
\end{equation}
running between $1/2$ for $R_R\gg R_L$ and  $-1/2$ for $R_R\ll
R_L$.

The pumping circuit is operational even at a constant gate
voltage, when $V_g=V_0$. Suppose that the capacitor may be
typified by a parallel plate capacitor with a tunable area $A$,
separation $d$ and permittivity $\varepsilon$. Initially the
capacitor is at equilibrium with charge $Q_0=C
V_0=\frac{\varepsilon A V_0}{2 \pi d}$. Then its area is varied to
$A^\prime=A-\delta A$ while $\alpha \approx -1/2 \; (R_R \ll
R_L)$. As a result a charge $ \delta Q_0 $ will leave the
capacitor to the right until equilibrium is restored. Then by
changing the area back from $A^\prime$ to $A$ when $\alpha \approx
1/2$, we will pull the same amount of charge $\delta Q_0$ into the
capacitor from the left. Repeating this process with a period
$\tau$ will result in an average pumped current $\delta C V_0
/\tau$ flowing from left to right.\cite{Rect03}~(A periodic
variation of $d$ or $\varepsilon$ will have a similar effect.)

Proceeding formally, let $Q(t)$ be the instantaneous charge on the
capacitor. The charge leaving the capacitor during a small time
interval $dt$ is $-d Q(t)=-\dot Q(t) dt$. From current
conservation and Kirchhoff`s rules:
\begin{equation}
I_L+I_R=  \dot {Q}(t) \equiv I_c,\qquad I_L R_L=I_R R_R,
\end{equation}
the fraction of this out going charge, $-d Q(t)$, leaving via
$R_L$ or via $R_R$ is
\begin{equation}
\label{eq:currentpartition}
dQ_{R(L)}(t)=-\frac{R_{L(R)}}{R_L+R_R}d Q(t).
\end{equation}
We define the pumped current as $I =\frac{\langle I_L \rangle
-\langle I_R \rangle}{2}$ where $\langle O \rangle=
\tau^{-1}\int_0^\tau O(t) dt$ denotes average over a period. Then
$I$ may be expressed in terms of $Q(t)$ as
\begin{equation}
\label{eq:a} I =\frac{1}{\tau}\int_0 ^\tau dt
\frac{1}{2}(I_L-I_R)= \frac{1}{\tau}\int_0 ^\tau dt \alpha (t)
\dot Q (t).
\end{equation}

The charge $Q(t)$ should be determined by the equations of motion
of the system which are governed by a Lagrangian $\cal{L}$,
including a source term in the Euler-Lagrange equations which
introduces dissipation:
\begin{equation}
\label{Euler} \frac{\delta  {\cal{L}}  (Q,\dot Q)}{\delta
Q}-\frac{d}{d t} \frac{ \delta {\cal{L}}(Q,\dot Q)}{\delta \dot
Q}=R \dot Q,
\end{equation}
where $R=R_L\parallel R_R=R_L R_R/(R_L+R_R)$.

If the parameters, $x(t),y(t),\dots$, that control $Q$ and
$\alpha$ in Eq.~(\ref{eq:a}) are varied slow enough then $Q$ and
$\alpha$ are functions of the instantaneous value of these
parameters, and do not depend explicitly on time:
\begin{eqnarray}
\label{eq:adiabatic_limit}
Q(t)&=&Q\left[x(t),y(t),\dots \right], \nonumber \\
\alpha(t)&=&\alpha\left[x(t),y(t),\dots \right].
\end{eqnarray}
The parameters $x(t),y(t),\dots$ can be for example $V_g, d, A,
\varepsilon, R_L,R_R$ or any combination of them, e.g., $\alpha$
itself. We will refer to this slow limit as the {\em adiabatic}
limit. For each case that we study we will check how large should
be the period $\tau$ for the adiabatic limit to be established.
Roughly, the adiabatic condition is established when $\tau$ is
larger than the effective $RC$ time of the circuit.

\begin{figure}
\begin{center}
\includegraphics*[width=80mm]{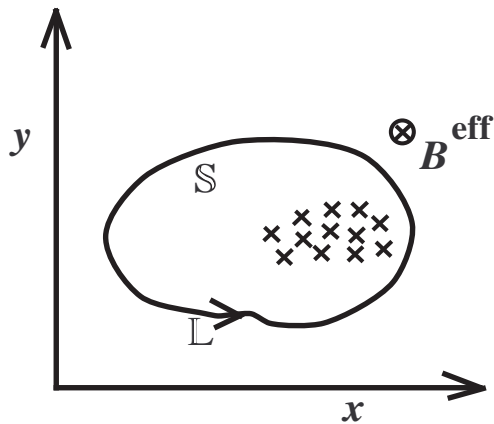}
\caption{The pumped charge per period is the flux of an effective
magnetic field inside the parameters trajectory.} \label{fg:beff}
\end{center}
\end{figure}

If there are only two parameters, then in the adiabatic limit the
pumped current is
\begin{equation}
I=\frac{1}{\tau}\int_0 ^\tau dt \alpha [x(t),y(t)] \dot Q
[x(t),y(t)].
\end{equation}
Using now
$$\frac{d Q}{dt} =\frac{\partial Q}{\partial x}  \frac{d
x}{dt} +\frac{ \partial Q}{\partial y} \frac{d
 y}{dt},$$
the current can be rewritten as a line integral along a trajectory
${\mathbb{L}}$ in the parameter space (see Fig.~\ref{fg:beff}):
\begin{equation}
I=\frac{1}{\tau}\oint_{{\mathbb{L}}} (\alpha \partial_x Q,\alpha
\partial_y Q)\cdot(dx,dy).
\end{equation}
Because the parameters are varied periodically in time the
trajectory is closed. Using Stoke's theorem, the line integral can
be transformed into a surface integral on the closed surface
$\mathbb{S}$ bounded by the trajectory ${\mathbb{L}}$,
\begin{eqnarray}
\label{eq:Beff} I&=&\frac{1}{\tau}\int\!\!\!\!\int_{\mathbb{S}}
dxdy
B^{\rm{eff}} = \frac{\phi^{\rm eff}}{\tau}, \nonumber \\
\nabla \times A^{\rm eff} &=&
B^{\rm{eff}}=\partial_x\alpha\partial_y Q-\partial_y \alpha
\partial_x Q = \nabla \alpha \times \nabla Q, \nonumber \\
A^{\rm eff} &=& \alpha \nabla Q \text{ or }\frac{1}{2}\left(\alpha
\nabla Q -Q \nabla \alpha \right),
\end{eqnarray}
where $B^{\rm{eff}}$ is an effective magnetic field in parameter
space. The last ambiguity in the definition of $A^{\rm eff}$ is a
result of a gauge freedom: an addition of the gradient term
$\frac{1}{2} \nabla( \alpha Q)$ does not change $B^{\rm eff}$.

Equation~(\ref{eq:Beff}) suggests that the charge pumped per
cycle, $\tau
 I$, is equal to the flux  of the effective magnetic field through
the loop in parameter space $\phi^{\rm eff}$,  as depicted in
Fig.~\ref{fg:beff}. Similar topological formulation for the pumped
current was discussed in the quantum case \cite{Brouwer98,Avron02}
while here we obtained a similar structure for the classical
situation.

\subsection{Example: $x=\alpha$, $y=V_g$} \label{specialcasse}
Since no current is pumped if the asymmetry parameter $\alpha$ is
kept fixed, we consider the simplest example where $\alpha$ itself
is a pumping parameter. Substituting the Lagrangian of the circuit
depicted in Fig.~\ref{fg:circuit} (generalized to the case where
the capacitance depends on its charge or on the voltage across
it):
\begin{eqnarray}
\label{lagrangian} {\cal{L}}(Q,\dot Q)=-E_c(Q)+Q V_g, \nonumber \\
E_c(Q)=\int_0^{Q}dQ^\prime V_c(Q^\prime), \nonumber \\
V_c(Q)=\int_0^{Q}\frac{d Q ^\prime }{C(Q^\prime)}
\end{eqnarray}
in Eq.~(\ref{Euler}), one gets the equation of motion
$V_g-V_c(Q)=\dot Q R$. Using $C=dQ/dV_c$ we get
\begin{equation}
\label{eq:eom_simlpe_circuit} V_g - V_c=R C \dot V_c.
\end{equation}

The meaning of the adiabatic limit is clear in this equation. The
typical time scale for changes in $V_c$ is the period time $\tau$.
Therefore if $ RC \ll \tau$ we can neglect the right hand side of
the equation and establish the adiabatic limit $V_c(t)=V_g(t)$.
Substituting this into Eq.~(\ref{eq:Beff}) one gets
\begin{equation} \label{I}
I=\frac{1}{\tau}\int \!\!\!\! \int_\mathbb{S} C(V_g) d\alpha dV_g,
\end{equation}
so that $C$ plays the role of the effective magnetic field.

For $C=10^{-15}F$ at a frequency $\tau^{-1}=1GHz$ and with a gate
voltage oscillations of $1mV$ the maximal pumped current is $1
nA$.

\subsection{Application to a more general electrical circuit}

\label{se:pumpsinseries} 
\begin{figure}[h]
\begin{center}
\includegraphics*[width=80mm]{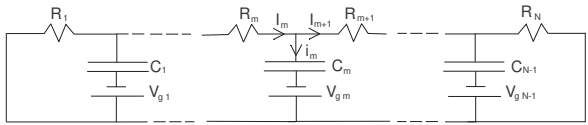}
\caption{Generalization of the circuit in
Fig.~\protect{\ref{fg:circuit}.}}\label{generalcircuit}
\end{center}
\end{figure}

Consider the electrical circuit shown in
Fig.~\ref{generalcircuit} that generalizes the circuit depicted in
Fig.~\ref{fg:circuit}. The analysis of the two circuits is
similar: let us assume that we change the charge on capacitor
$C_{m_0}$ by $d Q_{m_0}$ (by changing $V_{g~m_0}$), while keeping
the charge on the other capacitors constant. Using Kirchhoff's
rules one can show that the fraction of $d Q_{m_0}$  flowing to
the left via $R_1$, $d Q_L$, or to the right via $R_N$, $d Q_R$,
is:
\begin{equation}
\label{eq:gencurrentpartition} d Q_{R(L)}(t)=-
\frac{R_{m_0}^{L(R)}}{\sum_{m=1}^N R_m}d Q_{m_0}(t),
\end{equation}
where $R^{L}_{m_0}=\sum_{m=1}^{m_0} R_m$ and
$R^{R}_{m_0}=\sum_{m=m_0+1}^N R_m$ are the resistances to the left
and to the right of capacitor $C_{m_0}$ respectively.

The superposition principle generalizes
Eq.~(\ref{eq:gencurrentpartition}) to
\begin{equation}
\label{eq:gencurrentpartition1} d Q_{R(L)}(t)= - \frac{\sum_k
R_k^{L(R)}d Q_k(t)}{\sum_{m=1}^N R_m}
\end{equation}
for any variation of $\{ V_{g~k} \}$. Similarly, Eq.~(\ref{I})
which holds in the adiabatic limit becomes
\begin{equation} \label{II}
I=\frac{1}{\tau}\int \!\!\!\! \int_\mathbb{S} \sum_{k=1}^{N-1}
C(V_{g~k}) d\alpha_k dV_{g~k},
\end{equation}
where $\alpha_k=\frac{1}{2}\frac{R^{R}_k-R^L_k}{R^R_k+R^L_k}$.

In the next section we will show that a small quantum system
subjected to dephasing can be described by the circuit depicted in
Fig. \ref{generalcircuit}.

\section{Connection with Quantum Pumps subjected to Dephasing}
 \label{se:comparison}
The scattering approach for pumps\cite{Brouwer98} relates the DC
current flowing through a time dependent scatterer with its
scattering matrix and with the time derivatives of the scattering
matrix. However, in real physical systems there are processes
that lead to uncertainty in the phase of the single electron; for
example interaction with phonons and interaction between the
electrons mediated by the electromagnetic environment. This effect
is called dephasing. It is expected that a classical description
will gradually take place as the dephasing becomes stronger.

 To study the effect of dephasing on the quantum coherent pump in a
controllable way we introduce dephasing
leads~\cite{Buttiker86a,Brouwer02} and show that in the limit of
strong dephasing the pumping current approaches the classical
limit given in Eq.~(\ref{II}). The effect of dephasing can be
described in different ways which differ in \emph{details}, but
this variety of ways does not alter the main conclusion that a
quantum system with strong dephasing can be described by a
classical theory.

\begin{figure}[h]
\begin{center}
\includegraphics*[width=80mm,height=100mm]{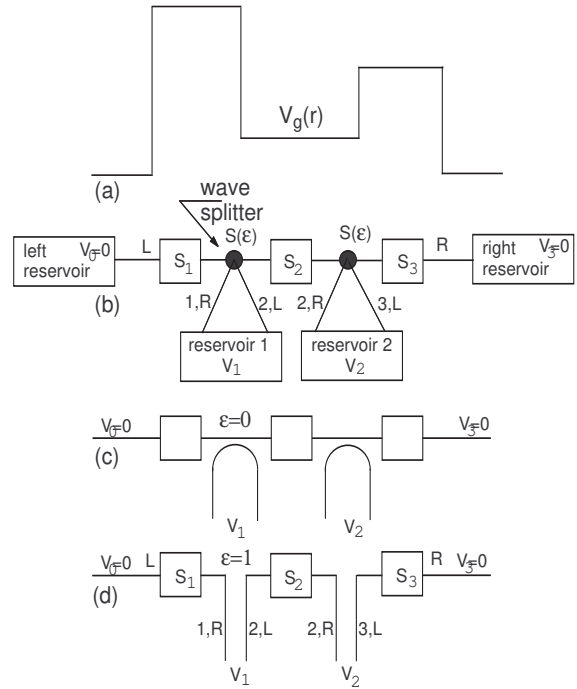}
\caption{(a) The gate potential along the original quantum system
without dephasing. (b) Introduction of dephasors, controlled by a
parameter $\epsilon$. (c) $\epsilon=0$: The system is totaly
coherent. (d) $\epsilon=1$: This configuration doesn't allow any
interference effects between different scatterers
$S_m$.}\label{dephasing}
\end{center}
\end{figure}

Consider a conducting wire subjected to a gate potential $V_g(r)$,
$r \in [0,L]$, as shown in Fig. \ref{dephasing}(a). To introduce
dephasing in a controlled way we connect $N-1$ wave splitters of 4
legs \cite{Footnotefork03} at the points $r=i \ell$, $i =
{1,2,...,N-1}$ along the wire, as shown in Fig. \ref{dephasing}(b)
(for $N=3$). The length $\ell$ determines the dephasing length of
the model together with the wave splitter parameter $\epsilon$, as
we will explain later. The wave splitters are described by the
scattering matrix
\begin{equation*}
\textbf{S}_{\rm{splitter}}(\epsilon)=\left(%
\begin{array}{cccc}
  0 & \sqrt{1-\epsilon} & \sqrt{\epsilon} & 0 \\
  \sqrt{1-\epsilon} & 0 & 0 & \sqrt{\epsilon} \\
  \sqrt{\epsilon} & 0 & 0 & -\sqrt{1-\epsilon} \\
  0 & \sqrt{\epsilon} & -\sqrt{1-\epsilon} & 0 \\
\end{array}%
\right),
\end{equation*}
where the third and forth lines and columns of the matrix
correspond to the two legs of the reservoir that serves as a
dephasor. Each wave splitter $i$ is connected to a reservoir $i$
held at voltage $V_i$. To mimic a dephasor that influences only
the phase coherence of the waves in the sample and does not
influence the total current, we tune the voltages $\{ V_i \}$ so
that \emph{no net charge flows into any of the reservoirs}. In our
time dependent problem we will assume that such conditions hold at
all times.

(In a similar model of dephasing~\cite{Das03} one introduces an
extra phase $\phi$ in selected points, which is averaged out
\emph{after} squaring the desired amplitude. The approach that we
use differs from the later in the following way: the introduction
of the reservoirs is accompanied by the addition of contact
resistances, which renders the total resistance from left to right
higher.)

For general $\epsilon$ the whole system, including the dephasing
leads, is described by a \textbf{S} matrix of rank $2N$. The
current in each lead is composed of two contributions, the pumped
current (Brouwer formula~\cite{Brouwer98}) and the biased current
(Landauer-Buttiker-Imry conductance formula~\cite{Landauer57}).
The voltages $\{ V_i \}$ produce biased currents which cancel the
pumped currents in each reservoir.

In the limit $\epsilon\rightarrow 0$ (without dephaing) the
\textbf{S} matrix is block diagonal: it is a direct sum of $N$
rank $2$ matrices. One of them connects between the left and right
reservoirs and describes scattering from the potential $V_g(r)$
between $r=0$ and $r=L$. The other matrices trivially connect each
reservoir only to itself, i.e., the reservoirs  are not connected
to the wire, see Fig.~\ref{dephasing}(c).

In the limit $\epsilon\rightarrow 1$ we again have
$\textbf{S}=\oplus_{m=1}^{N} \textbf{S}_m$. Now $\textbf{S}_m$
describes a potential barrier $V_{g(m)}(r)$ between two dephasors
given by
\begin{equation}
\label{eq:step} V_{g(m)}(r) = \left \{
\begin{array}{ll}
V_g(r) & \textrm{$(m-1) \ell<r<m \ell$}  \\
0 & \textrm{else}
\end{array} \right. .
\end{equation}

The original barrier is split into $N$ sections, see
Fig.~\ref{dephasing}(d). Notice that while the phase coherence
between different sections is lost, they are correlated by the
requirement for zero current in the reservoirs. The dephasing
length is thus equal to $\ell$.

Let us concentrate on the $\epsilon= 1$ case which is easier to
handle, (mentioning that, in principle the model could be solved
for any $\epsilon$) and parameterize the unitary scattering matrix
of each section as
\begin{equation}
\label{eq:Sm} \textbf{S}_m=e^{i \Lambda_m}
\left(%
\begin{array}{cc}
  e^{i \eta_m}\cos \theta_m & i e^{-i \phi_m} \sin\theta_m \\
  i e^{i \phi_m} \sin\theta_m & e^{-i \eta_m}\cos\theta_m \\
\end{array}%
\right).
\end{equation}
For this parametrization the Brouwer formula reads \cite{Avron02}
\begin{eqnarray} \label{fridel}
2 \pi/ e \; dQ_{m,L}=- d\Lambda_m-\cos ^2 \theta_m
d\eta_m+ \sin ^2 \theta_m d\phi_m \nonumber\\
2 \pi/ e \; dQ_{m,R}=- d\Lambda_m+\cos ^2 \theta_m d\eta_m- \sin
^2 \theta_m d\phi_m,
\end{eqnarray}
where $dQ_{m,L}$ and $dQ_{m,R}$ are the pumped charges flowing out
of section $m$ to the left and to the right respectively, see
Fig.~\ref{dephasing}.

To understand how pumping takes place in the $\epsilon= 1$ case
suppose first we change $V_g(r)$ only in section $m_0$ such that
only the parameters $\Lambda_{m_0},\theta_{m_0},\eta_{m_0}$ and
$\phi_{m_0}$ may change. Consequently there are pumped charges
flowing out of (or into) section $m_0$ to the left and to the
right in leads $m_0,L$ and $m_0,R$, given by Eq.~(\ref{fridel}).
The change in the charge of section $m_0$ (by analogy with the
charge of the capacitor in Sec.~{\ref{se:Elec_circ}}), is
\begin{equation}
\label{totcharge} dQ_{m_0}=-(dQ_{m_0,L}+dQ_{m_0,R}).
\end{equation}
We denote its partitioning into left and right by
$-(1/2+\alpha^Q_{m_0})\; dQ_{m_0} \equiv dQ_{m_0,L} $ and
$-(1/2-\alpha^Q_{m_0})\; dQ_{m_0} \equiv dQ_{m_0,R}$, where
$\alpha_{m_0}^Q$ is called the quantum partitioning coefficient of
section $m_0$. (When $d Q_{m_0}=0$ one should write the formulae
below explicitly in terms of $dQ_{m_0,L}$ and $dQ_{m_0,R}$.)

The voltages on the reservoirs $m = 1, \dots, N-1$ are adjusted to
cancel out the pumping contribution, so that there is zero current
in each reservoir. That gives the equations
\begin{eqnarray} \label{probe}
\frac{V_m-V_{m-1}}{ R^Q_{m}}+\frac{V_m-V_{m+1}}{R^Q_{m+1}}
 &= & \nonumber \\- \bigl( \delta_{m,m_0-1} (1/2+\alpha^Q_{m_0})&+&\delta_{m,m_0} (1/2-\alpha^Q_{m_0}) \bigr)\dot{Q}_{m_0}; \nonumber
 \\
m&=&1,2,\dots, N-1,
\end{eqnarray}
where $R_m^{Q}=(\frac{e^2}{h}\sin^2 \theta_m)^{-1}$ are the
quantum resistances of the different sections. The currents into
the left and the right reservoirs are
\begin{equation}
\label{eq:leftrightcurrents}
\begin{array}{cccccc}
d Q_L&=&V_1/ R^{Q}_1 dt&-& \delta_{m_0,1} &(1/2+\alpha^Q_{m_0}) \; d Q_{m_0},  \\
d Q_R&=&V_{N-1}/ R^{Q}_N  dt&-&\delta_{m_0,N}
&(1/2-\alpha^Q_{m_0}) \; d Q_{m_0}.
\end{array}
\end{equation}

Solving Eq.~(\ref{probe}) for $\{V_m\}_{m=1}^{N-1}$ we
find:\cite{Soldif03}

\begin{eqnarray}
\label{U} dQ_L & = &- \frac{(1/2+\alpha^Q_{m_0})
R^{Q}_{m_0}+\sum_{m=m_0+1}^{N}
R^{Q}_m}{\sum_{m=1}^{N} R^{Q}_m}dQ_{m_0}, \nonumber\\
dQ_R & = &- \frac{\sum_{m=1}^{m_0-1} R^{Q}_m+(1/2-\alpha^Q_{m_0})
R^{Q}_{m_0}}{\sum_{m=1}^{N} R^{Q}_m}dQ_{m_0}.
\end{eqnarray}

We thus obtain Eq.~(\ref{eq:currentpartition}) with $R_L
\rightarrow \sum_{m=1}^{m_0-1} R^{Q}_m+(1/2-\alpha^Q_{m_0})
R^{Q}_{m_0}$ and $R_R \rightarrow (1/2+\alpha^Q_{m_0})
R^{Q}_{m_0}+\sum_{m=m_0+1}^{N} R^{Q}_m$.

During a generic pumping process the potential is varied in
different places, i.e., we have to consider simultaneous
variations of $V_g(r)$ in sections $m \ne m_0$. Since the
different sections are connected classically when $\epsilon=1$,
and Eq.~(\ref{probe}) is linear in the source term $\propto \dot
Q_{m_0}$, we sum over all the contributions to $dQ_L$ and $dQ_R$
arising from each section. We then obtain
Eq.~(\ref{eq:gencurrentpartition1}) in which the resistors are
obtained from the Landauer resistances $\{ R^Q_m \}$ and the
quantum partitioning coefficients $\{ \alpha^Q_m \}$ as
\begin{equation}
R_m=(1/2+\alpha^Q_{m-1}) R^Q_{m-1}+(1/2-\alpha^Q_{m}) R^Q_m.
\end{equation}

First, notice that for large $N$ Eq.~(\ref{U}) is insensitive to
the actual value of $\{ \alpha^Q_m \}$. It occurs because a change
in these values modifies only one term in the sum. Second, as we
already mentioned above,  this mapping is impossible when
$dQ_{m_0,L}=-dQ_{m_0,R}$, i.e., when the effect of the variations
in the potential in a single coherent section is to transfer
charge from left to right, without charging nor discharging this
section. This situation occurs only for a specially tuned
asymmetric potential inside a single coherent section. However,
our equivalent circuit requires a nearly constant potential in
each section, as there is only one gate in each section (see
Fig.~\ref{generalcircuit}). Notice that for large $N$ this is a
small effect: one can see from Eq.~(\ref{U}) that for this case
$dQ_{L(R)}= \frac{R^{Q}_{m_0}}{\sum_{m=1}^{N}
R^{Q}_m}dQ_{m_0,L(R)}$.

\subsection{Comparison between coherent and non-coherent effects}
 \label{se:comp}

The dephasors connect \emph{ incoherently} (or \emph{classically})
sections of length $\ell$, which conserve internal phase
coherence. Indeed Eq.~(\ref{U}) is derived by classical circuit
theory, however, coherence effects still determine its parameters
($dQ_{m_0}$ and the resistances $\{R^Q_m\}$). In this section we
separate coherent from non-coherent contributions to $dQ_{m_0}$
and compare between their magnitudes.

To do so, suppose the gates length $l_{\rm{gate}} \gg \ell$ such
that the characteristic length scale for changes of $V_g(r)$ is
larger than~$\ell$. Then $V_g(r)$ can be approximated by
Eq.~(\ref{eq:step}) where $V_{g(m)}(r)$ is a \emph{rectangular}
barrier of width $\ell$ and height $V_{g(m)} \equiv V_g\bigl(
\ell(m-1/2)\bigr)$. The transmission coefficient of section $m$ is
thus
\begin{equation}
\label{t} t_m=\frac{4 k_F k_m e^{ i k_m
\ell}}{(k_F+k_m)^2-(k_F-k_m)^2 e^{2 i k_m \ell}},
\end{equation}
where $k_F$ is the Fermi wave number and $k_m=\sqrt {2m
(E_F-V_{g(m)})/\hbar^2} $ is the wave number inside the barrier of
section $m$. (Since all the potential voltages $\propto \omega$,
for very small $\omega$ the Fermi wave numbers in all sections are
similar and we ignore the difference between them.)

Consider one of the sections that are influenced by the long gate,
say $m_0$. Due to the inversion symmetry of each barrier we have
$\eta_{m_0}=0$ and $\phi_{m_0}=0$, and thus
$\Lambda_{m_0}=\arg(t_{m_0})$. As $V_{g(m_0)}$ increases,
particles are pushed out symmetrically (in this case
$\alpha_{m_0}^Q=0$), and by Eq.~(\ref{fridel}) their charge is
$dQ_{m_0,L}+dQ_{m_0,R}=-dQ_{m_0}=-e d \Lambda_{m_0}/ \pi$.
Eq.~(\ref{t}) can be derived by summing over all possible wave
trajectories from the left to the right of the barrier; the
numerator corresponds to the straight (classical) path through the
barrier while the denominator appears after summing the rest of
the paths. Following this observation we artificially decompose
$dQ_{m_0}$ into a non-coherent part [numerator of Eq.~(\ref{t})]
and coherent part [denominator of Eq.~(\ref{t})]:
\begin{equation}
\label{eq:aasw} \frac{dQ_{m_0}}{e dV_{g(m_0)}}  =
-\theta(E_F-V_{g(m_0)})  D_{m_0} \ell  -\frac{ d \arg f_{m_0}}{\pi
d V_{g(m_0)}},
\end{equation} where $D_{m_0}=\frac{m}{ \pi \hbar^2 k_{m_0}}$ is the density of states at
wave number $k_{m_0}$ and $f_{m_0}$ is the denominator of
$t_{m_0}$ in Eq.~(\ref{t}). We want to compare these two terms for
a finite change in gate voltage $V_{g(m_0)}$.

Since $|k_F+k_m|\geq|k_F-k_m|$ it is straight forward to realize
that $\arg f_{m_0}$ is an oscillatory function of $V_{g(m_0)}$
whose amplitude of oscillations is smaller than $\pi$. Therefore
the contribution of the  second term in Eq.~(\ref{eq:aasw}), a
consequence of interference of many paths, is bounded by the
single electron charge, $e$, for any variation of $V_{g(m_0)}$.

Now we have to sum over the quantum contributions of the $l_{\rm
gate} /\ell$ sections that are influenced by the gate. Due to the
oscillatory nature of this term, if the gate length
$l_{\rm{gate}}\gg \ell$ and we assume the oscillations to be
uncorrelated then the contribution of the coherent term is
$\mathcal{O}( \sqrt{\l_{\rm{gate}} / \ell} e)$.

On the other hand the first classical term, which is simply the
number of states below the Fermi level in a 1D box of size $\ell$
with potential $V_{g(m_0)}$, is not bounded. Furthermore the
contributions due to various sections of length $\ell$ which are
subjected to the influence of the same gate potential add up and
give charge $\mathcal{O}( l_{\rm gate}/\ell\; \tilde q e)$ where
$\tilde q$ is the average number of states added to a section of
length $\ell $ due to the change in the gate potential.

We see that the non-coherent (classical) effect dominates when
$l_{\rm gate} \gg \ell$, or when $\tilde q \gg 1$. In that
situation coherent effects are unimportant and our electrical
treatment of pumping is relevant; the pumped current is then given
by minus the expression in Eq.~(\ref{II}) with $C(V) = \frac{e
\ell \sqrt{m} }{\pi \hbar \sqrt{2 (E_F-e V)}} \theta(E_F-e V)$.

On the other hand, the classical term can be tuned to zero if we
oscillate several gates together, such that no net charge flows
into or out of the system, and then the current results only from
quantum interference. This happens for example when two nearby
gates change in opposite directions, or in the two dimensional
case when the gates change the \emph{shape} of the capacitor but
not its \emph{area}.

To conclude, a quantum pump with dephasors can be described by the
electrical circuit of Fig.~\ref{generalcircuit} where each section
of the circuit describes a coherent section of length $\ell$ of
the quantum system. The components of the circuit (resistors,
capacitors, etc.) are generally determined by coherent effects
(for example a capacitor $C_i$ might depend on the voltage
$V_{g~i}$ in an oscillatory way). In contrast to coherent systems,
for the classical circuit (i) the charge inside a coherent section
is determined only by the potential in that region,
Eqs.~(\ref{fridel}) and ~(\ref{totcharge}); (ii) the left-right
partitioning of an extra charge flowing out of (or into) a
coherent section is simply the classical partition of current
through two parallel resistors, Eq.~(\ref{U}); (iii) the
superposition principle holds for arbitrary change of the
parameters in different sections.

\subsection{Example}

To illustrate the difference between the classical and quantum
aspects of pumping consider as an example the potential
$V(r)=\gamma\delta(r)+U\theta[r(L-r)]$. For a square shaped path
of the parameters $(\gamma,U)$ passing through the points $(0,0)$,
$(\infty,0)$, $(\infty,\delta U)$ and $(0,\delta U)$, the pumped
current is \cite{Brouwer98}
\begin{eqnarray}
\label{delta} I&=&\frac{e  L \omega \delta U}{8 \pi^2 k_F}+
\frac{e\omega \delta U}{16 \pi^2 k_F^2} \bigl( \pi
\sin^2(k_FL)-\sin(2 k_F L) \bigr)+ \nonumber
\\  &+&\mathcal{O}(\delta U^2)
\end{eqnarray}
with $\frac{\hbar^2}{2m}=1$. Let us add $N-1$ dephasing leads as
described above and tune $\epsilon=1$. The matrix $\textbf{S}_1$
describes the original barrier shrinked to length $\ell$ instead
of $L$ while the matrices $\textbf{S}_m$, $m=2,...,N$ describe
rectangular barriers of height $U$ and width $\ell$. To find the
current in the presence of dephasing we have to follow the 4
stages of the period using Eq.~(\ref{U}) and Eq.~(\ref{fridel}).
For example, in the third part, where the parameters $(\gamma,U)$
go from $(\infty,\delta U)$ to $(0,\delta U)$, only the phases
$\Lambda_1$, $\theta_1$, and $\eta_1$ change. To first order in
$\delta U$, the resistances of the $N-1$ barriers are equal to the
quantum resistance, and thus
\begin{eqnarray}
\label{111} \frac{\delta Q_R}{e}&=&\int_0 ^\infty \frac{\sec^2
\theta_1}{\sec^2 \theta_1 +N-1}  \nonumber \\  &\times& (\frac{
d\Lambda_1}{ d \gamma}-\cos^2 \theta_1 \frac{ d\eta_1}{  d
\gamma}) \frac{d\gamma}{4 \pi},
\end{eqnarray}
where $\theta, \eta,\Lambda$ and $\eta$ are defined according to
Eq.~(\ref{eq:Sm}). Performing similar calculations for the rest of
the period, one obtains a DC current,
\begin{eqnarray}
\label{delta1} I&=&\frac{e L \omega  \delta U}{8 \pi^2
k_F}+\frac{e\omega \delta U}{16 \pi^2 k_F^2} \bigl( \frac{\pi
\sin^2(k_F \ell)}{\sqrt N}-\sin(2 k_F \ell)(2-\frac{1}{N})
\bigr)\nonumber
\\  &+&\mathcal{O}(\delta U^2),
\end{eqnarray}
directed to the right.

Let us compare Eq.~(\ref{delta}) with Eq.~(\ref{delta1}). The
first terms are identical, independently of $N$. These terms are
the classical contribution which is unaffected by the dephasing,
being related only to local density of states. The second terms
coincide for $N=1$ but differ otherwise. The factor $\sin(2kL)$ in
Eq.~(\ref{delta}) was transformed into $\sin(2k \ell)$, i.e.,
coherent effects are restricted to the dephasing length. [This
result depends on the dephasors position: a single dephasor at
$r=0^+$ is enough to destroy the interference term completely,
since to $\mathcal{O}(\delta U)$ the only section giving rise to
interference effects is the one with the $\delta$-function.]

The ratio between the classical and the interference terms is of
the order of $L /\lambda_F$, confirming the importance of the
classical term when $L \gg \lambda_F$ (or $l_{\rm{gate}} \gg
\lambda_F$).

There are both interference terms $\mathcal{O}(N^0)$ which survive
the limit $N \rightarrow \infty$, and others,
$\mathcal{O}(N^{-1})$ and $\mathcal{O}(N^{-1/2})$, which disappear
in that limit. The reason for having interference effects in the
large $N$ limit in this example is the presence of the infinite
barrier: when the parameters $(\gamma,U)$ go from $(\infty,0)$ to
$(\infty,\delta U)$, all the charge repelled from sections
$m=1,...,N-1$ is driven to the right. Among all these sections,
only section $m=1$ gives rise to the interference term which is
independent of $N$. On the other hand when the parameters move
from $(\infty,\delta U)$ to $(0,\delta U)$, $\delta Q_R$ is given
by Eq.~(\ref{111}) and we see that for $N=\infty$ we get $I_R=0$.

\section{Spin Polarized DC Current}
\label{se:spin pola}

In Sec.~\ref{se:Elec_circ} we analyzed the time evolution of an
electrical circuit and derived an expression for the pumping of
charge in the adiabatic limit. In Sec.~\ref{se:comparison} we used
a dephasing model to show that in the limit of strong dephasing
the non interacting coherent pumping expression is reduced to the
classical expression. In this section we will generalize the
classical equations to include spin. We will assume (i) the
dephasing length is smaller than the size of the system, $\ell \ll
L$, so that the classical expressions are valid; (ii) the spin is
conserved during a pumping cycle, i.e., $\tau \ll \tau_{\rm{sf}}$,
with $\tau_{\rm{sf}}$ being the mean spin-flip time.

The generalization of the spinless treatment of
Sec.~\ref{se:Elec_circ} to the spinfull case is done by
introducing a Lagrangian that depends on charges with spin up and
spin down. The equations of motion with the dissipation term are
[cf. Eq.~(\ref{Euler})]
\begin{equation}
\label{Eulerspin} \frac{\delta {\cal{L}}(Q_\sigma,\dot
Q_\sigma)}{\delta Q_\sigma}-\frac{d}{d t} \frac{ \delta
{\cal{L}}(Q_\sigma,\dot Q_\sigma)}{\delta \dot Q_\sigma}=R \dot
Q_\sigma
\end{equation}
where we have assumed that $R$ doesn't depend on spin. Now the
current and voltages have two components, referring to spin up and
down. The analysis goes in parallel to the spinless case: given
the time dependent charge with spin index $\sigma$ on the
capacitor, $Q_\sigma(t)$, the pumped current is
\begin{equation}
I_\sigma= \frac{1}{\tau}\int dt \alpha (t) \dot Q_\sigma (t).
\end{equation}
In the adiabatic limit the spinfull form of Eq.~(\ref{eq:Beff}) is
\begin{eqnarray}
\label{Beffspin}I_\sigma&=&\frac{1}{\tau}\int\!\!\!\!\int_{\mathbb{S}}
d \vec S \cdot \vec B^{\rm{eff}}_\sigma,\qquad \vec
B^{\rm{eff}}_{\sigma}=\vec \nabla \alpha \times \vec \nabla
Q_\sigma,
\nonumber \\
I_{c,s}&=&\frac{1}{\tau}\int\!\!\!\!\int_{\mathbb{S}} d \vec S
\cdot \vec B^{\rm{eff}}_{c,s}, \qquad \vec B^{\rm{eff}}_{c,s}=\vec
B^{\rm{eff}}_\uparrow\pm \vec B^{\rm{eff}}_\downarrow,
\end{eqnarray}
where $I_{c,s}=I_\uparrow \pm I_\downarrow$. The 2 dimensional
integrals are done on the area $\mathbb{S}$ bounded by the
trajectory $\mathbb{L}$ in the parameter space. The $\vec \nabla$
symbol denotes partial derivative with respect to the parameters
$(x,y,z)$, $\vec \nabla=(\partial_x,\partial_y,\partial_z)$. The
increase of the number of parameters from 2 to 3, cf.
Eq.~(\ref{eq:Beff}), reflects the need to control the spin
polarization of the pumped current.

\begin{figure}[h]
\begin{center}
\includegraphics*[width=50mm]{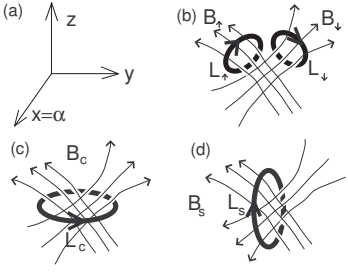}
\caption{(a) Parameter space. (b) Illustrative effective magnetic
field lines for spin up and down. The loops
${\mathbb{L}}_\uparrow$ and ${\mathbb{L}}_\downarrow$ correspond
to pumping of spin up and down electrons, respectively, from left
to right. (c) The field lines for charge pumping are obtained by
adding the spin up and spin down field lines. The loop
${\mathbb{L}}_c$ corresponds to pumping of charge, i.e.,
unpolarized electrons. (d) The effective magnetic lines for spin.
The loop ${\mathbb{L}}_s$ winds around the spin lines but, in
total, it does not wind around the charge lines. Thus it
corresponds to pure spin pumping without charge
transfer.}\label{spincharge}
\end{center}
\end{figure}

To understand better the meaning of the effective magnetic field
$\vec B^{\rm{eff}}_\sigma$ let us decompose it into field lines,
as shown schematically in Fig. \ref{spincharge}. According to Eq.
(\ref{Beffspin}) the total charge of spin $\sigma$ pumped from
left to right per cycle, associated with a given loop
${\mathbb{L}}$, is the flux of field lines of spin $\sigma$ inside
the loop ${\mathbb{L}}$. To illustrate this point, the loops
${\mathbb{L}}_\uparrow$ and ${\mathbb{L}}_\downarrow$ in Fig.
\ref{spincharge}(b) correspond to pumping of only spin up and spin
down, respectively.

In Figs. \ref{spincharge}(c) and \ref{spincharge}(d) we show the
charge and spin field lines obtained by adding or subtracting the
spin up and down field lines drawn in Fig. \ref{spincharge}(b).
Loop ${\mathbb{L}}_c$ winds around charge lines but in total it
does not wind around spin lines and thus it corresponds to pumping
of unpolarized charge from left to right. In contrast, loop
${\mathbb{L}}_s$ winds around spin magnetic lines only and
corresponds to pumping of pure spin. By pumping spin we mean that
electrons with opposite spins are transported in opposite
directions through the pump. Putting all field lines together we
reconstruct $\vec B^{\rm{eff}}_\sigma$. Assuming it is continuous,
the conditions for the existence of a finite spin current without
any charge transport (for infinitesimal $\mathbb{S}$) are $ d \vec
S \cdot \vec B^{\rm{eff}}_{s} \neq 0 $ and $d \vec S \perp \vec
B^{\rm{eff}}_{c}$.

Specializing to the case $x=\alpha$, the effective magnetic field
in Eq.~(\ref{Beffspin}) becomes
\begin{equation}
\label{Beffspin1} \vec B^{\rm{eff}}_{\sigma}=(0,-\partial_z
Q_\sigma,\partial_y Q_\sigma).
\end{equation}
The fact that $\vec B^{\rm{eff}} \perp x$ ensures that there is no
pumping if $\alpha$ is kept constant.

We have completed our general classical analysis of the spin
pumps. In the next section we will apply Eqs.~(\ref{Beffspin}) and
(\ref{Beffspin1}) to a particular system of experimental interest.

\section{Application to Two Dimensional Electron Gas}
\label{se:2DEG} In the preceding sections we have performed a
general classical analysis of spin and charge pumps. We have
argued that with sufficiently strong dephasing our analysis is
valid. In this section we will apply the general theory
[Eqs.~(\ref{Beffspin}) and (\ref{Beffspin1})] to a 2DEG and
propose a realization of a \emph{spin
battery}.\cite{Coulombblockade03} The pumping parameters will be
$x=\alpha$ determining the asymmetry between the contacts
connecting the 2DEG to the left and right leads, $y=V_g e$ being a
plunger gate and $z=h=g\mu_B B$ being the Zeeman energy associated
with an in-plane magnetic field.

To find $\vec B^{\rm{eff}}_\sigma$ we have to calculate the
dependence of the charge of the 2DEG on the pumping parameters,
$\vec \nabla Q_\sigma$ [see Eq.~(\ref{Beffspin})]. The density of
particles of each spin in the system, $n_\sigma$, is determined by
the grand canonical ensemble average, and thus it is a function of
the chemical potentials and of the parameters $\vec
r=(x,y,z,...)$:
$n_\sigma=n_\sigma(\mu_\uparrow,\mu_\downarrow,\vec r)$ or
$\mu_\sigma=\mu_\sigma(n_\uparrow,n_\downarrow,\vec r)$. In the 2D
case $Q_\sigma=n_\sigma e A$ where $A$ is the 2DEG area. The
differentials $d \vec r$, $dn_\sigma$ for which the system remains
at equilibrium satisfy $d\mu_\sigma=\sum_{\sigma^\prime}\partial
\mu_\sigma/\partial n_{\sigma^\prime} |_{n_{\bar{\sigma}^\prime }
,\vec r}\cdot dn_{\sigma^\prime}+\vec \nabla \mu_\sigma
|_{n_\uparrow,n_\downarrow}\cdot d \vec {r} =0$. This leads to the
equality
\begin{equation}
\label{dqdx} \vec \nabla Q_\sigma \equiv \frac{\partial
Q_\sigma}{\partial x_i}=-e  A \sum_{\sigma^\prime}D_{\sigma
\sigma^\prime} \vec \nabla
\mu_{\sigma^\prime}|_{n_\uparrow,n_\downarrow},
\end{equation}
where the thermodynamical density of states tensor (DOS) was
introduced,
\begin{equation}
\label{DDOOSS} D_{\sigma\sigma^\prime}\equiv\frac{\partial
n_\sigma}{\partial \mu_{\sigma^\prime}}.
\end{equation}
The quantity $(\nabla \mu_{\sigma})_i$ is referred as {\em the
$r_i$ inverse compressibility of spin $\sigma$}.

When the energy is given explicitly as function of $n_\sigma$ and
$r_i$ we can obtain the chemical potential, the $r_i$ inverse
compressibility of spin $\sigma$ and the DOS tensor:
\begin{eqnarray}
\label{partial}
\mu_\sigma & = & \partial_{n_\sigma} E(n_\uparrow,n_\downarrow,\vec r), \nonumber \\
 (\nabla \mu_{\sigma}|_{n_\uparrow,n_\downarrow})_i  & = &
\partial^2_{n_\sigma,r_i}
E(n_\uparrow,n_\downarrow,\vec r),  \nonumber  \\
F_{\sigma,\sigma^\prime}&=&\frac{\partial\mu_\sigma }{\partial
n_{\sigma^\prime}};\qquad \textbf{D}  = \textbf{F}^{-1}.
\end{eqnarray}

Assuming that the pumping parameters influence the energy of the
system only via the terms $-e(n_\uparrow+
n_\downarrow)V_g-h(n_\uparrow - n_\downarrow)$ then from
Eq.~(\ref{partial}) it follows $(\nabla
\mu_{\sigma}|_{n_\uparrow,n_\downarrow})_y=-1$ and $(\nabla
\mu_{\sigma}|_{n_\uparrow,n_\downarrow})_z=-\sigma$ with
$\sigma=\pm 1$. Using
$D_{\downarrow\uparrow}=D_{\uparrow\downarrow}$ and
Eq.~(\ref{Beffspin}) we get
\begin{eqnarray} \label{vecC}
\vec{B}^{\rm{eff}}_\sigma & = & eA \bigl( 0,\sigma(D_{\sigma\overline{\sigma}}-D_{\sigma\sigma}),D_{\sigma\sigma}+D_{\sigma\overline{\sigma}} \bigr) \nonumber   \\
\vec{B}^{\rm{eff}}_c & = &
eA(0,D_{\downarrow\downarrow}-D_{\uparrow\uparrow},D_{\uparrow\uparrow}+D_{\downarrow\downarrow}+2D_{\uparrow\downarrow})
\nonumber \\ \vec{B}^{\rm{eff}}_s & = &
eA(0,2D_{\uparrow\downarrow}-D_{\uparrow\uparrow}-D_{\downarrow\downarrow},D_{\uparrow\uparrow}-D_{\downarrow\downarrow}),
\end{eqnarray}
with $x=\alpha$, $y=e V_g$, $z=h$ and $\overline{\sigma}=-\sigma$.
Together with Eq.~(\ref{Beffspin}) we can express the pumped
current, $I_{\sigma}$, in terms of the DOS tensor.

There are some interesting relations concerning the DOS tensor: in
a pumping process in which $h$ (Zeeman energy) remains
\emph{constant} we find that only the third component of $\vec
B^{\rm{eff}}$ contributes. Using Eq.~(\ref{vecC}), the ratio
between the spin and charge currents in this case is
\begin{equation}
\label{eq:IsIcpump}
\frac{I_s}{I_c}=\frac{B^{\rm{eff}}_{s,z}}{B^{\rm{eff}}_{c,z}}=\frac{D_{\uparrow\uparrow}-D_{\downarrow\downarrow}}
{D_{\uparrow\uparrow}+D_{\downarrow\downarrow}+2D_{\uparrow\downarrow}}.
\end{equation}
In equilibrium we have $\mu_\uparrow=\mu_\downarrow \equiv \mu$.
MacDonald showed \cite{MacDonald99} that the inverse magnetic
compressibility, $\frac{\partial\mu}{\partial
h}|_{n_\uparrow+n_\downarrow}$, is given by minus the expression
in  Eq. (\ref{eq:IsIcpump}).\\
Similarly, \emph{at constant $V_g$} one can verify that
$\frac{I_s}{I_c}=\frac{\partial\mu}{\partial
h}|_{n_\uparrow-n_\downarrow}$. Therefore measurement of charge
and spin pumping currents reveals thermodynamical properties of
the system.

Now we will evaluate the spin and charge pumped currents in a 2DEG
relying on the Hartree-Fock approximation for its
energy:~\cite{Stern73}
\begin{eqnarray}
\label{HF}  E  (n_\uparrow,n_\downarrow,y,z) &=&
\frac{n^2_\uparrow+n^2_\downarrow}{2D_0}+\frac{e^2(n_\uparrow+n_\downarrow)^2}{2
\tilde{C}}-{}\nonumber\\-\frac{8e^2(n^{3/2}_\uparrow+n^{3/2}_\downarrow)}{3\sqrt{\pi}}&-&
y(n_\uparrow+ n_\downarrow)-z(n_\uparrow - n_\downarrow),
\end{eqnarray}
where $D_0^{-1}=2 \pi \hbar ^2 / m^*$. The first term is the
kinetic energy, the second and third terms are the Hartree and the
exchange interactions respectively, the forth term describes
interaction with an external gate potential $e V_g=y$ and the last
term describes interaction with an external magnetic field $h=g
\mu_B B=z$. As we will see, the negative sign of the exchange term
increases the magnetic susceptibility  at low densities.

\subsection{Non-interacting electrons}
\label{free} To discuss the noninteracting case we disregard the
terms $\propto e^2$ in Eq.~(\ref{HF}).
One finds then using Eq.~(\ref{partial}) that $\textbf{D}=\left(%
\begin{array}{cc}
  D_0 & 0 \\
  0 & D_0 \\
\end{array}%
\right)$. Equation~(\ref{vecC}) gives the effective magnetic
fields, in units of $eAD_0$ (charge per unit energy),
\begin{equation}
\label{befffree} \vec{B}^{\rm{eff}}_\sigma  = (0,-\sigma,1) \qquad
\vec{B}^{\rm{eff}}_c  = (0,0,2)  \qquad \vec{B}^{\rm{eff}}_s  =
(0,-2,0).
\end{equation}
 The
facts that $\vec{B}^{\rm{eff}}_c\|z$ and $\vec{B}^{\rm{eff}}_s\|y$
($y=h$, $z=V_g$) mean that a small loop pointing in the $z$
direction (changing $\alpha$ and $V_g$) produces only charge
current, and a small loop pointing in the $y$ direction (changing
$\alpha$ and $h$) produces a pure spin current. (The direction of
a infinitesimal loop is normal to the plane of the loop.)

To estimate the pumped currents for oscillating $\alpha$ and $V_g$
or magnetic field $B$ we use the expression $I_{c,s}=2 \tau^{-1} e
AD_0 \delta r_{c,s} \delta \alpha$ derived from
Eqs.~(\ref{Beffspin}) and (\ref{befffree}). Here the energy
$\delta r_c$ is $\delta V_g e$, when only charge is pumped, and
$\delta r_s$ is  $\delta h$, when only spin is pumped.

If  we assume that the left-right resistors of the pump oscillate
with maximal amplitude, $\delta \alpha =1$, and we consider a GaAs
sample (g=-0.44) of area $A=1 \mu m^2$, then the charge current
obtained for $\delta V_g=3mV$ and frequency $\tau^{-1}=10 GHz$ is
$1.3 \mu A$. At high frequencies it is difficult experimentally to
produce oscillatory magnetic field with a sizable Zeeman energy.
Thus, for a magnetic field of order $\delta B=1mT$ and frequency
$\tau^{-1}=10 KHz$ the spin pumped current is very small $\sim
.5~10^{-15} A$.

\subsection{Capacitive interaction} \label{se:CoulombIn} The
simplest way to consider the Coulomb interaction is performed by
adding a capacitive term to the Hamiltonian. Thus, adding the
second term in Eq.~(\ref{HF}) one has in units of $eAD_0$
\begin{eqnarray}
\vec{B}^{\rm{eff}}_\sigma & = &(0,-\sigma,1/(1+2D_0e^2/\tilde{C} ))  \nonumber \\
\vec{B}^{\rm{eff}}_c & = &(0,0,2/(1+2D_0e^2/\tilde{C})) \nonumber \\
\vec{B}^{\rm{eff}}_s & = &(0,-2,0).
\end{eqnarray}
The observation that the magnitude of $\vec{B}^{\rm{eff} }_c$ has
decreased reflects the fact that the energy needed to add charge
to the system is now larger. On the other hand, $\vec{B}^{\rm{eff}
}_s$ and hence the spin current are unaffected by the capacitance
term.

The capacitance between the 2DEG and the gate separated one from
the other by a distance $d$ is $C=\frac{\varepsilon A}{2 \pi d}$.
Then, the reduction factor $1+2D_0e^2/\tilde{C}$ can be written as
$1+\frac{2d}{a_0^{\rm{eff}}}$ where $a_0^{\rm{eff}}$ is the
effective Bohr radius $\approx 100$\AA~ in GaAs. For a separation
of $d=1000$\AA~ the reduction factor is $20$. An estimate for the
pumped charge in the presence of the capacitance with $\delta
V_g=30mV$, under the conditions specified above, yields $I_c
\simeq 0.66 \mu A$.

In contrast to biased transport, where the conductivity is
proportional to the aspect ratio in two dimensions, the current of
the pump is proportional to the area $A$ as can be seen in the
pre-factor of $\vec B^{\rm{eff}}$. There are two reasons to bound
the area of the pump from above. One is that spin-flip events may
spoil the spin polarization in too big samples. This restricts the
typical lengths to $\sim \mu m$.\cite{Prinz98} The other
restriction comes from the adiabatic condition, which for the
simple circuit of Fig.~(\ref{fg:circuit}) reads $R C \ll \tau$,
where the capacitance is proportional to the area $A$. Rewriting
the energy per unit area in terms of density, $n=n_\uparrow +
n_\downarrow$ and magnetization, $m=n_\uparrow - n_\downarrow$, as
\begin{eqnarray}
\label{Enm} E(n,m)&=&\left[m-m_0(h)\right]^2 \frac{1}{4 D_0} \nonumber \\
&+& \left[n-n_0(V_g)\right]^2 \left( \frac{1}{4 D_0}+ \frac{e^2}{2
\tilde C}\right),
\end{eqnarray}
where $n_0$ and $m_0$ are the average density and magnetization
respectively, we can read off the effective capacitances
corresponding to charge, $C_c=(\frac{1}{C}+\frac{1}{2 A D_0
e^2})^{-1}$, and to spin, $C_s=2AD_0e^2$, where $C=\tilde{C}A$. We
see that $C_s > C_c$, thus the adiabatic condition $R \max \{
C_c,C_s \}  \ll \tau$ becomes $A \ll \frac{\pi \tau
a_0^{\rm{eff}}}{\varepsilon R}$. For the above mentioned
conditions with $R \equiv R_L
\parallel R_R=1K \Omega$, the adiabatic bound is $A \ll 400 \mu
m^2$.

\subsection{Exchange interaction} \label{se:hartree fock}

The next step is to include the exchange term, the third in
Eq.~(\ref{HF}). Now the DOS tensor depends on electron density:
$\textbf{F}=\textbf{D}^{-1}$ and
$F_{\sigma\sigma`}=\delta_{\sigma\sigma`}
D_0^{-1}+\frac{e^2}{\tilde{C}}-\delta_{\sigma\sigma`}\frac{2e^2n^{-1/2}_\sigma}{\sqrt{\pi}}$,
which is different from the noninteracting 2D systems, where the
DOS is constant. The dependence on density leads to a spontaneous
magnetization at low densities corresponding to $r_s \cong
2$.\cite{MacDonald99} The spontaneous magnetization occurs in our
model due to the following facts: (i) at constant density
$n=n_\uparrow+n_\downarrow$ the exchange term is minimized for
maximal $|m|$, i.e., for full polarization; (ii) in sufficiently
low densities the exchange term dominates, since it behaves as
$n^{3/2}$ and the others as $n^2$.

Before the spontaneous magnetization occurs application of
magnetic field strongly polarizes the spins $(n_{\uparrow} \neq
n_{\downarrow})$ and hence $D_{\uparrow \uparrow} - D_{\downarrow
\downarrow} \propto n^{-1/2}_\uparrow-n^{-1/2}_\downarrow \neq 0$.
This imbalance between the spin up and spin down populations at
weak magnetic field becomes stronger at low densities. Using
Eq.~(\ref{vecC}) we see that except for special points in
parameter space it is not true that $\vec{B}^{\rm{eff}}_c \| z$
and $\vec{B}^{\rm{eff}}_s\|y$. From the last fact it follows that
one can achieve spin current without varying magnetic field. This
spin current is proportional to $D_{\uparrow \uparrow} -
D_{\downarrow \downarrow}$ and becomes larger at low densities.

\begin{figure}[h]
\begin{center}
\includegraphics*[width=70mm]{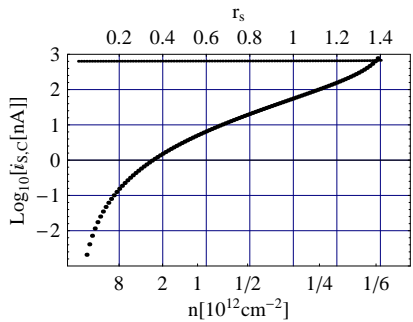}
\caption{Spin current (big dots) ($\propto
D_{\uparrow\uparrow}-D_{\downarrow\downarrow}$) and charge current
(small dots) ($\propto
D_{\uparrow\uparrow}+D_{\downarrow\downarrow}+2D_{\uparrow\downarrow}$)
as function of density in a 2DEG according to the Hartree Fock
approximation.}\label{fighartree}
\end{center}
\end{figure}

Following the last observation we calculated $I_s$ and $I_c$ as
function of $r_s=\frac{1}{a_0^{\rm{eff}}\sqrt{\pi n}}$ at a
\emph{constant magnetic field} of $1T$. The conditions were
similar to those specified in Sec.~\ref{free} with $d=1000$\AA,
$\delta V_g=30mV$ and $\delta \alpha=1$. The results are plotted
in Fig. \ref{fighartree}. The charge current is almost invariant
$I_c \simeq 0.66 \mu A$. On the other hand, as explained in the
previous paragraph, $I_s$ grows rapidly as the density is
decreased until the ground state becomes ferromagnetic at $r_s
\approx 1.4$ (the critical $r_s$ is now smaller due to the
magnetic field).

We do not expect the Hartree-Fock approximation to be reliable for
$r_s>1$. Monte Carlo calculations \cite{Tanatar89} show that, in
reality this instability does not occur until much lower densities
are reached. For $r_s=3/4$ we have $I_s=15.6 nA$. It should be
emphasized that the existence of $I_s$ is a consequence of the
interactions between electrons in two dimensions. Notice  that
this spin current flows in the background of a large charge
current of $0.64 \mu A$ and may be hard to detect.

This difficulty can be resolved by a manipulation of the puming
contour, using the fact that though the spin current is small,
i.e., $B^{\rm{eff}}_s \ll B^{\rm{eff}}_c$, it is rapidly varying
at low densities, $\frac{d}{dV_g} B^{\rm{eff}}_s \gg
\frac{d}{dV_g} B^{\rm{eff}}_c$. (In Fig.~\ref{fighartree} one can
see how these quantities depend on density.) According to our
topological interpretation, a small loop in parameter space probes
the effective magnetic field lines passing through it. In order to
probe the \emph{derivative} of $B^{\rm{eff}}$, and not its
magnitude, we will use a figure $8$ shaped trajectory whose
equation in the $(\alpha,V_g)$ plane is $\alpha(t)=\delta \alpha
\sin(2 \omega t)$, $V_g(t)=\delta V_g \sin(\omega t)$. Basically,
this trajectory is composed of two adjacent loops with
\emph{opposite} directions. Applying such a pumping trajectory for
the case of the 2DEG, means the following: in the first half of
the period (one loop) we pump a charge with small polarization, as
we found in our previous calculation at $r_s=3/4$. In the second
half we pump a nearly equal amount of charge to the opposite
direction, since the direction of the second loop is reversed. In
the second part, however, the average density is different, thus
the spin polarization is different. The excess spins may lead to a
pure spin current.

The calculation for the pumped currents with the figure $8$ shaped
loop, as defined above with $\delta \alpha = 1/2$ and $\delta
V_g=50mV$ yields $I_c=1.2nA$, $I_s=5.2nA$, $\frac{I_s}{I_c}=4.3$.
In order to cancel the residual $1.2nA$ charge current we will
consider an \emph{asymmetric} figure $8$ shaped loop. This can be
done using the parametrization in the $(\alpha,V_g)$ plane given
above with an additional term in $V_g$ equal to $\varepsilon
\delta V_g |\sin(\omega t)|$, where the asymmetry of the
trajectory is determined by the parameter $|\varepsilon|<1$. We
find that $\frac{I_s}{I_c}$ diverges at $\varepsilon_0 \cong
-10^{-3}$. However, getting exactly to the point where the charge
current is cancelled requires a fine tuning of $\varepsilon$. For
example, to obtain $|\frac{I_s}{I_c}|
> 40$, the parameter $\varepsilon$ should be close to $\varepsilon_0$
as $|\varepsilon-\varepsilon_0|< 2 \times 10^{-4}$. This means
that a resolution of about $10 \mu V$ in the gates voltage is
required.

\subsection{Temperature effects}

We would like to estimate the temperature needed for the operation
of the spin pump discussed in Sec.~\ref{se:hartree fock}. Consider
the quadratic part of Eq.~(\ref{HF}), given in Eq.~(\ref{Enm}).
The thermal smearing of the magnetization $m$ is equal to $\Delta
m_T \simeq \sqrt{\frac{ D_0 K_B T}{A}}$ (The presence of the
exchange term does not change $\Delta m_T$ significantly). In the
pumping process the parameters oscillate and change the
magnetization $m$ during the period with an amplitude $\Delta
m_P$, which is related to the pumped spin current by $\Delta m_P
\simeq \frac{I_s \tau}{A e}$. We impose that the thermal smearing
of $m$ is small relative to the parameters induced oscillations in
$m$, i.e., $\Delta m_T \ll \Delta m_P$. This yields $ T \ll
(\frac{I_s \tau}{e })^{2} \frac{1}{K_B D_0 A}\cong 10K $ where
$I_s=5.2nA$ was used. Finally, the thermal smearing of the density
$n$ is smaller than $\Delta m_T$ by the factor
$\sqrt{\frac{C_c}{C_s}} \cong 4.4$ for the above specified
conditions.

\section{Summary}
\label{se:Summary} We analyzed classical non-coherent pumps and
discussed their connection with quantum pumps using a dephasing
model. Our classical method can include electron-electron
interactions and allows to study the effect of exchange
interaction on spin pumping. We expressed the classical pumped
spin current in terms of the thermodynamic DOS tensor of the
system, and gave a topological interpretation to it in terms of
effective ``magnetic" flux through trajectory loops in the space
of the parameters that control the pump [see Eq.~(\ref{eq:Beff})].
We analyzed in detail the case of 2DEG GaAs and found that any
combination of charge and spin currents can be obtained by
choosing appropriate trajectory in parameter space. In particular
one can choose a trajectory that corresponds to a pure spin
current, which has magnitude of order of nano-Ampers.

\section{acknowledgement}
\label{se:acknowledgement}

We are grateful to Alessandro Silva for many discussions at the
initial stages of the research, and  to Ora Entin-Wolman, Amnon
Aharony, Yehoshua Levinson, Yoseph Imry, Charles Marcus, Felix von
Oppen and Jens Koch for useful comments. Special thanks to David
J. Thouless for the enlighten remark concerning the topological
interpretation of the spin pump.

This work was supported by  Minerva, by German Israeli DIP C 7.1
grant and by the Israeli Science Foundation via Grant No.160/01-1.

\appendix

\section{Pumped currents Vs. Bias currents}
\label{se:pumpvsbiased}

According to Einstein's relation the current of spin
$\sigma=\uparrow,\downarrow$ is
\begin{equation}
\label{eq:Einstien}
 I_{\rm{bias},\sigma}/  V_{\rm
 bias}=e^2
 D_{\mathrm{if}}  \frac{\partial n_\sigma}{\partial \mu} ~\times~[\rm{aspect~ratio}],
\end{equation}
where ${D_{\mathrm{if}}}$ is the diffusion constant. Using the
definition of the DOS tensor, Eq. (\ref{DDOOSS}), we find that in
equilibrium ($\mu_\uparrow=\mu_\downarrow \equiv \mu$) we have
$\frac{\partial n_\sigma}{\partial \mu}=\sum_{\sigma'}D_{\sigma
\sigma'}$.

Assuming that the diffusion constant is spin independent we find,
[similar to Eq.~(\ref{eq:IsIcpump}) for the case of pumps]
\begin{equation}
\label{eq:IsIcbias} \frac{I_{\rm bias}^s}{I_{\rm bias}^{c}} \equiv
\frac{I_{\rm{bias},\uparrow}-I_{\rm{bias},\downarrow}}{I_{\rm{bias},\uparrow}+I_{\rm{bias},\downarrow}}
=
\frac{D_{\uparrow\uparrow}-D_{\downarrow\downarrow}}{D_{\uparrow\uparrow}+D_{\downarrow\downarrow}+2D_{\downarrow\uparrow}}.
\end{equation}
This shows that bias currents and classical pumped currents have
similar expressions in terms of the DOS tensor, cf
Eq.(\ref{vecC}). Now we will show that a pure spin current, on
which we elaborated in Sec.~\ref{se:hartree fock} for the case of
pumping, can be obtained by an oscillatory bias together with an
in-phase oscillating gate.

Consider an AC biased 2DEG in a constant magnetic field, where a
gate $V_g$ oscillates in phase with the bias:
$V_{\rm{bias}}(t)=V_b \sin(\omega t)$, $V_{\rm{gate}}(t)=V_g
\sin(\omega t)$. Due to the oscillating gate a rectified DC
component in the current will appear. Using Einstein's relation,
this DC current is given by
\begin{eqnarray}
\label{recti} \langle I_c\rangle &=&e^2
 D_{\mathrm{if}}
\langle(D_{\uparrow\uparrow}+D_{\downarrow\downarrow}+2D_{\downarrow\uparrow})(t)V_{\rm{bias}}(t)\rangle
~\times~[\rm{a.~r.}]
\nonumber \\
\langle I_s\rangle &=&e^2
 D_{\mathrm{if}}
\langle(D_{\uparrow\uparrow}-D_{\downarrow\downarrow})(t)V_{\rm{bias}}(t)\rangle
~\times~[\rm{a.~r.}]
\end{eqnarray}
where $\langle  \rangle$ denotes time average over a period and we
assumed that the diffusion constant does not depend on $V_{\rm
bias}$ or $V_{\rm gate}$.

Near the magnetic instability the spin current $\propto
D_{\uparrow\uparrow}-D_{\downarrow\downarrow}$ is very susceptible
to changes in $V_g$ while the charge current $\propto
D_{\uparrow\uparrow}+D_{\downarrow\downarrow}+2D_{\downarrow\uparrow}$
is nearly constant (see Fig.~\ref{fighartree}). As a result, the
charge flowing in one half of the period cancels the charge
flowing in the second half, while a finite amount of spins
 is transferred from left to tight.

This rectification~\cite{Rect03} effect is based on the same
mechanism as the adiabatic pumping effect considered in
Sec.~\ref{se:hartree fock}. Both effects are proportional to the
interaction strength.

We will show now that the two effects of pumping and rectification
(bias) produce currents of the same order, provided that (i) the
pumping frequency saturates the adiabatic condition, $\tau=R C$
and (ii) the pumping gate, the rectification gate and the bias
oscillate with similar amplitudes.

First consider a simple pumping loop in which the asymmetry
parameter $\alpha$ and the pumping gate oscillate out of phase
with $\delta \alpha=1$, $\delta V_g=V$. When the adiabatic
condition is saturated, $\tau=R C$, the pumped current in the
circuit of Fig.~\ref{fg:circuit}, given in Eq.~(\ref{I}), is
$I_{\rm{pump}}=\tau^{-1} C V \delta \alpha =V /R$. Using Ohm`s law
we see that the DC current flowing due to a DC bias $V$ in the
same circuit is of the same order, i.e., $I_{\rm{bias}}=V/2R$,
where we assumed $R \equiv R_L
\parallel R_R \simeq R_L /2 \simeq R_R /2$ for simplicity.

Second, let us compare the spin currents for pumping with a figure
8 shaped loop and for rectification (i.e., bias voltage) {\em with
an oscillating gate}. We can apply the above argument for each
half period of either pumping or rectification processes: for
pumping, each half period consists of one of the two loops of the
figure 8. In each such loop the amplitude of the gate oscillation
is half of the oscillation in the entire period. For the case of
rectification, in each half period the averaged bias is of the
order of the maximal AC amplitude. In both cases, the density
varies between the two halves of the periods: the voltage
providing different densities corresponds to (i) the distance
between the centers of the two loops in figure 8 for the case of
pumping; (ii) the oscillations of the gate in the case of
rectification. Thus when the pumping gate, the rectification gate
and the bias oscillate with similar amplitudes similar spin
currents will be created in the pumping and in the rectification
processes.

To conclude, unless biasing of the system is unwanted, one can
obtain spin current either by the pumping effect discussed in
Sec.~\ref{se:hartree fock} or by the rectification effect combined
with an oscillating gate described here.

On the other hand, as is seen in Appendix \ref{se:linearresponse},
for long systems adiabatic pumping is more feasible than
application of bias, since it requires only small voltages.

\section{Connection with non-linear response theory}
\label{se:linearresponse}

In this section we compare our results to other theories
\cite{KamenevOreg95,Oppen01} that evaluate non linear response to
an external potential at wave number $k$ and frequency $\omega$.

We start by analyzing the circuit of Fig.~\ref{generalcircuit} in
the continuous limit, with the assumptions (i) $R_m=R(V_{g~m})$
and $C_m=C(V_{g~m})$ so that the varying parameters are
$\{V_{g~m}\}$; (ii) periodic boundary conditions: $R_1$ is
connected to $R_N$.

The continuous version of the current conservation in the
junctions of Fig.~\ref{generalcircuit} is, in the adiabatic limit,
\begin{equation}
\label{discrete1} \frac{dI(r,t)}{dr}=-\tilde{C}
\bigl(V_g(r,t)\bigr)\dot{V}_g(r,t), \quad \omega \tilde R \tilde C
k^{-2} \ll 1,
\end{equation}
where $I(r)$ is the current along the circuit and
$\tilde{C},\tilde R$ are the capacitance and resistance per unit
length respectively. To estimate the adiabatic condition we use
the characteristic length scale $k^{-1}$. The periodic boundary
conditions imply
\begin{equation}
\label{kirjov}\int_0^l dr I(r,t)\tilde{R}\bigl(V_g(r,t)\bigr)=0,
\end{equation}
where $l$ is the length of the system. Integrating
Eq.~(\ref{discrete1}) and substituting into Eq.~(\ref{kirjov}) we
have:
\begin{eqnarray}
\label{aaa}I(0,t)l \tilde {R}( \bar V_g(t)) &=& \int_0^ldr  \tilde{R}\bigl(V_g(r,t)\bigr)\nonumber \\
&\times&\int_0^r dr'\tilde{C}\bigl(V_g(r',t)\bigr)\dot{V_g}(r',t),
\end{eqnarray}
where $\bar {V_g}(t)=l^{-1}\int_0^ldr V_g(r,t)$.

 We proceed by assuming that the potential $V_g(r,t)$ depends on two
parameters $x,y$ as
\begin{equation}
V_g(r,t)=x(t)\sin(kr)+y(t)\cos(kr),
\end{equation}
and take $\tilde {C}\bigl(V_g(r,t)\bigr)=\tilde {C}$ to be a
constant. The integral over $r'$ in Eq.~(\ref{aaa}) yields
\begin{equation}
\frac{\tilde{C}}{k} \left[\dot x(t) \bigl(1- \cos(kr)\bigr)+\dot
y(t)\sin(kr) \right],
\end{equation}
and the integral over $r$ yields
\begin{equation}
\label{eq:eq} I(0,t)=\frac{\tilde{C}\tilde{R}'}{2k
\tilde{R}}\bigl(x(t) \dot{y}(t)-y(t) \dot{x}(t)\bigr)+\frac{\tilde
C}{k} \dot{x},
\end{equation}
where we expanded $\tilde R$ around $\bar V_g=0$ as $\tilde R
(V_g)=\tilde R+ \tilde{R}' V_g$. Integrating Eq.(\ref{eq:eq}) over
one period as done in Eq.~(\ref{eq:Beff}), we get the charge per
period for small area $\mathbb{S}$:
\begin{eqnarray}
\label{eq:adaada} Q_p=\frac{\tilde{C}\tilde{R}'}{k
\tilde{R}}\int_{\mathbb{S}} dx dy  \Rightarrow  \nonumber \\
j_{\rm DC}= \frac{\tilde{R}' \tilde{C}}{
\tilde{R}}\frac{\omega}{k} V^2.
\end{eqnarray}
in the last equality we have assumed that $x(t)=\sqrt{2} V \cos
(\omega t)$, $y(t)=\sqrt{2} V \sin (\omega t)$.

We can compare our results to the non liner response theory.
Ignoring coherent effects,\cite{KamenevOreg95} the current
response to a time and space dependent potential $V(r,t)$ is
$j(r,t)=-\sigma[n(r,t)] \nabla V(r,t)$. If $V(r,t)$ induces a
density polarization given by $\delta n(k,\omega)=-\Pi(k,\omega) e
V(k,\omega)$, the  DC response current is given by
\cite{KamenevOreg95,Oppen01}
\begin{equation}
\label{eq:linearresponse} j_{\rm{DC}}=\sum_{k,\omega}\frac{d
\sigma}{d e n} k \Im \Pi(k,\omega) |e V(k,\omega)|^2.
\end{equation}

We take now the diffusive polarization
operator\cite{KamenevOreg95} in the limit $D_{\rm if} k^2 \gg
\omega$, this gives $\Im \Pi(k,\omega) \rightarrow -\frac{\omega
}{ D_{\rm if}k^2} \frac{dn}{d\mu}$. Thus,
\begin{equation}
\begin{array}{ccccc}
j_{\rm{DC}}&=&-\frac{d \sigma}{d \mu} \frac{d \mu}{d e n}
\frac{1}{D_{\rm if}} & e^2 \frac{d  n}{d \mu}& \frac{\omega}{k}V^2 \\
&=& \frac{\tilde R'}{\tilde R}  & \tilde C &  \frac{\omega}{k}
V^2,
\end{array}
\end{equation}
which is identical to Eq.~(\ref{eq:adaada}). Since $D_{\rm
if}=\frac{ \sigma}{e^2dn/\mu}=1/(\tilde R \tilde C)$ the limit
$D_{\rm if} k^2 \gg \omega$ is identical to the adiabatic limit,
Eq.~(\ref{discrete1}).


\begin{thebibliography}{31}
\bibitem[{Foo()}]{Footnotefork03}
\bibinfo{note}{We use the 4 leg version for controlled dephasing instead of the
  3 leg model since this enables us to obtain perfect dephasing. (see remark in
  \protect{\cite{Buttiker86a}}).}
\bibitem[{Sol()}]{Soldif03}
\bibinfo{note}{Consider for example the case $m_0 \ne 1,N$. In that case since
  $V_0=V_N=0$ and no pumping current flows through reservoirs $m \ne m_0,m_0-1$
  we have $V_{m_0-1}=V_1/R^Q_1 \sum_{m=1}^{m_0-1} R^Q_m,V_{m_0}=V_{N-1}/R^Q_{N}
  \sum_{m=m_0+1}^{N} R^Q_m$ and have to solve two simple linear equations for
  $V_1$ and $V_{N-1}$, that together with Eq.~(\ref{eq:leftrightcurrents}) give
  Eq.~(\ref{U})}.
\bibitem[{Cou()}]{Coulombblockade03}
\bibinfo{note}{Treatment of Coulomb-blockade situatiuons will take place in a
  future work \cite{Sela04}.}
\end{thebibliography}

\end{document}